\documentclass[prl,superscriptaddress]{revtex4}

\usepackage{graphicx}        

\begin{document}

\title{Correlated band structure of electron-doped cuprate materials}


\author{C. Dahnken}
\affiliation{Institute for Theoretical Physics and Astrophysics,
  University of W\"urzburg, Am Hubland, 97074 W\"urzburg, Germany}
\author{M. Potthoff}
\affiliation{Institute for Theoretical Physics and Astrophysics,
  University of W\"urzburg, Am Hubland, 97074 W\"urzburg, Germany}
\author{E. Arrigoni}
\affiliation{Institute for Theoretical Physics and Computational Physics,
Graz University of Technology, A-8010 Graz, Austria}
\author{W. Hanke}
\affiliation{Institute for Theoretical Physics and Astrophysics,
  University of W\"urzburg, am Hubland, 97074 W\"urzburg, Germany}
\affiliation{Kavli Institute for Theoretical Physics,
  University of California, Santa Barbara, California 93106-4030 }

%
%


\begin{abstract}
 We present a numerical study of the doping dependence of the spectral
 function of the n-type cuprates.  Using a variational
 cluster-perturbation theory approach based upon the
 self-energy-functional theory, the spectral function of the
 electron-doped two-dimensional Hubbard model is calculated. The model
 includes the next-nearest neighbor electronic hopping amplitude $t'$
 and a fixed on-site interaction $U=8t$ at half filling and doping
 levels ranging from $x=0.077$ to $x=0.20$.  Our results support the
 fact that a comprehensive description of the single-particle spectrum
 of electron-doped cuprates requires a proper treatment of strong
 electronic correlations.  In contrast to previous weak-coupling
 approaches, we obtain a consistent description of the ARPES
 experiments {\it without} the need to introduce a
 doping-dependent on-site interaction $U$.
%
\end{abstract}

\maketitle


\section{Introduction}
\label{sec:introduction}

Angular resolved photoemission spectroscopy (ARPES) has greatly contributed to
our current understanding of materials with strong electron correlations, in
particular high-temperature superconductors (HTSC).
Up to a few years ago, ARPES investigations have been concentrated on
hole-doped HTSC materials.
%
Since ARPES  probes the 
 part of the spectral function that is occupied by electrons, 
only the region below the insulating gap
can be investigated  in hole-doped cuprates.
Although the observation of the unoccupied parts of the spectral function is
in principle possible via inverse photoemission, the process is highly involved
and does not yield the same resolution as direct photoemission.
An opportunity for a more comprehensive study of the doping dependence of the
spectral function is offered by electron-doped cuprates.
In these materials,
not only the excitations below the Fermi level in the \emph{lower} Hubbard band,
but also those below the Fermi level in the \emph{upper} Hubbard band can be
studied and
thus a large part of the important low-energy excitations is covered.
For this reason,
the investigation of such n-type cuprates by ARPES provides a large
amount of new information on the quasiparticle dynamics of HTSC cuprates.
%
%

%
Recently, an ARPES study of the doping dependence of the electron-doped
cuprate Nd$_{2-x}$Ce$_x$CuOCl$_{4 \pm \delta}$ (NCCO) has been carried
out by the Stanford group~\cite{ar.lu.01,ar.ro.02}.
In these measurements,
the low-energy excitations of Nd$_{2}$CuOCl$_{4 \pm \delta}$ (NCO) were shown
to essentially coincide with 
the ones of typical  undoped parent compounds of hole-doped materials
such as
 Sr$_{2}$CuO$_2$Cl$_2$  and Ca$_{2}$CuO$_2$Cl$_2$,
thus demonstrating
the universality of the electronic structure of the (single layer)
undoped cuprates.
Upon electron doping, a remarkable  Fermi surface (FS) evolution was found:
In the heavily  underdoped region the low-energy spectral weight
is limited to an area close around $\mathbf{k}=(\pi,0)$.
This has been interpreted as the formation of electron pockets.
With increasing doping level, these  pockets are connected by FS patches and
finally form a large LDA-like FS closed around 
$\mathbf{k}=(\pi,\pi)$.

The spectral function and FS data presented in \cite{ar.ro.02} gave rise to
several theoretical interpretations, which also include the idea of a 
collapse of the Mott
gap due to  strong suppression of the local Coulomb repulsion upon doping\cite{ku.ma.02}.
This conclusion was based upon mean-field calculations which employ a
fitting
 of the on-site repulsion $U_\mathrm{eff}$ of the
Hubbard model as a function of doping, 
to the value of the antiferromagnetic order parameter obtained from
experiments at each doping level.

In the present paper, 
using both standard cluster-perturbation theory (CPT)
\cite{gr.va.93,se.pe.02,se.pe.00} and, in particular, also
a variationally improved version 
 (V-CPT) 
\cite{po.ai.03,da.ai.04,se.la.05,ai.ar.05u},
which is based on the self-energy functional theory~\cite{pott.03},
we calculate the spectral function of the two-dimensional one-band 
Hubbard model. This model is taken with next-nearest
neighbor hopping amplitude $t'=-0.35t$ and fixed on-site
interaction $U=8t$ at half filling and doping levels ranging from $x=0.077$
to $x=0.20$.
%
Our numerical results
show that the salient features of the recent ARPES
experiments for electron-doped cuprates can be reproduced with
one-and-the-same Hubbard model without the necessity to resort to
any  change of the $U$-values as a function of doping 
as used in previous theoretical studies.
Our challenge here is to reproduce the global (i.e. n- and p-doped) phase
diagram by one universal choice of the model parameters, starting
from a picture of a doped Mott-Hubbard insulator.
\section{Model}
\label{sec:model}


As a generic model for the
 HTSC compounds we use the one-band Hubbard
model \cite{hubb.63} $H_{1b}$, i.e.
\begin{equation}
    \label{eq:h1b}
    H_{1b}=-t \sum_{\left< i,j\right>} \left(c^\dagger_{i\sigma}c_{j\sigma} +
    h.c \right) 
  + U \sum_{i} n_{i\uparrow} n_{i\downarrow}.
\end{equation}
Here, $c^\dagger_{i\sigma}$ ($c_{i\sigma}$) creates (annihilates) an electron
on site $i$ with spin $\sigma$, $\left< ... \right>$ denotes nearest neighbors
and $U$ is the on-site part of the Coulomb repulsion.
%
%
Although the $t-U$ Hubbard model at low temperature develops a quasiparticle
band of the appropriate width \cite{gr.ed.00,gr.99}, the dispersion shows a
near degeneracy between the $\mathbf{k}=(\pi,0)$ and the
$\mathbf{k}=(\pi/2,\pi/2)$ points.
From ARPES experiments, however, we know that the quasiparticle peak at
$\mathbf{k}=(\pi,0)$ is shifted to higher binding energies.
Actually, the dispersion of the quasiparticle peak shows two parabola with
lowest binding energy at $\mathbf{k}=(\pi/2,0)$ and $\mathbf{k}=(\pi,\pi/2)$.
It is, thus, indispensable to add at least one additional
term~\cite{to.ma.00,du.na.97u},  which  is taken here 
to be the hopping between next-nearest
neighbors ($\left<\left< ... \right>\right>$), i.e.
\begin{equation}
  \label{eq:h1bnn}
    -t' \sum_{\left<\left< i,j\right>\right>} \left(c^\dagger_{i\sigma}c_{j\sigma} +
    h.c \right).
\end{equation}
Even longer-range hopping ($t''$) elements have been proposed to achieve
consistency with experiment \cite{se.tr.04}.  However, for the purpose of
a qualitative analysis, it is sufficient to lift the degeneracy between
$\mathbf{k}=(\pi,0)$ and $\mathbf{k}=(\pi/2,\pi/2)$ and, thus, create the
indirect gap as observed in experiments.


\section{Numerical Technique}
\label{sec:numerical-technique}

Despite the considerable simplification arising from the use of an effective
single-band model, the calculation of the spectral function of the Hubbard
model still remains a difficult task.
The calculation of this quantity by exact diagonalisation is only
possible for a small lattice of
up to $4 \times 4$
sites, provided periodic boundary conditions are used.
Larger lattice sizes can only be calculated by stochastic methods, such as
the quantum Monte Carlo (QMC) technique \cite{hirs.88} or the
density matrix renormalization group algorithm (DMRG) \cite{no.wh.94}.
While these techniques certainly represent powerful approaches to
strongly-correlated 
electron systems, they are known to show disadvantages when
applied to 
the present problem.
In case of QMC, doping and low temperatures lead to the well-known 
sign problem,
i.e. the computation time increases exponentially with $T$ and system size
\cite{gr.99,endr.th.96}.
DMRG, in contrast, is a ground state technique successfully applied
for 1-D and ladder systems, but
displays convergence problems when applied to two-dimensional systems.
%


Recently, a strong-coupling perturbation theory has been developed for
which the
infinite lattice is subdivided into sufficiently small clusters 
that 
 can be treated exactly, followed by an infinite-lattice expansion in
powers of the hopping between the clusters \cite{gr.va.93,se.pe.02,se.pe.00}.
The expansion in the intercluster hopping can be formally carried out up to
arbitrary order following the diagrammatic method of Refs.
\cite{metz.91,pa.se.00,pa.se.98}.
The lowest order of this strong-coupling expansion in the inter-cluster
hopping has been termed ``cluster perturbation theory'' (CPT).

The CPT Green's function is given by
\begin{equation}
  \label{eq:cpt1}
  \mathbf{G}_\infty=\mathbf{G}_0+\mathbf{G}_0 \mathbf{T} \mathbf{G}_0
  +\mathbf{G}_0 \mathbf{T} \mathbf{G}_0\mathbf{T} \mathbf{G}_0 \ldots 
= \left[ \mathbf{G}^{-1}_0 - \mathbf{T}\right]^{-1},
\end{equation}
where  $\mathbf{G}_\infty$ is the Green's function of the infinite
system, $\mathbf{G}_0$ the cluster Green's function and $\mathbf{T}$
the inter-cluster hopping. All quantities are matrices with 
indices referring to the particular cluster and to the sites within that
cluster.
CPT can be considered as a systematic approach with respect to the cluster
size, i.e. it becomes exact in the limit $N_c \rightarrow \infty$, where $N_c$
is the number of sites within a cluster.
In addition, CPT provides approximate results for an infinitely large
system:
One of the advantages of this method is that
the CPT Green's function is defined for any wave vector $\mathbf{k}$ in
the Brillouin zone, contrary to common ``direct'' cluster methods, like QMC or
ED, for which only a few momenta are available.

CPT results for static quantities as well as for the single-particle spectral
function have been shown to agree very well with different exact analytical
and numerical results \cite{se.pe.02,se.pe.00}.
On the other hand, there is also a serious disadvantage of  CPT at this
level: Namely, the method does not contain any self-consistent procedure which
implies that symmetry-broken phases cannot be studied.
We have recently proposed a variational approach (V-CPT) to this problem, which
is based on the self-energy-functional approach (SFA).
This method is explained in detail  elsewhere\cite{po.ai.03,pott.03,da.ai.04}.
We use V-CPT to calculate the Green's function of the half-filled Hubbard model
with long-range antiferromagnetic order while plain CPT is used for the doped 
system.
The extension to a doped system with $d$-wave superconductivity was
first carried out  in Ref.~\cite{se.la.05}, see also Ref.~\cite{ai.ar.05u}.

\section{Results and Discussion}
\label{sec:results}

We consider the above single-band Hubbard model with nearest- $t$ and
next-nearest-neighbor $t'$ hopping at zero temperature.
Useful parameterizations of the $t-t'-U$ Hubbard model can be taken from the
literature \cite{du.na.97u,bren.95}.
We choose $t'=-0.35t$ and $U=8t$ here, which yields a sufficiently accurate
ratio for the Mott gap $\Delta \approx 4t$ and for the width of the quasiparticle
band $W \approx 1t$, and fits the experimental dispersion of the quasiparticle
band.
For a reasonable approximation to the full many-body problem, CPT and V-CPT
calculations for relatively large clusters are required.
In this work, we have calculated the spectral function for half filling and for 
doped systems with $x=0.07$ to $0.2$.
To achieve the smallest doping level, clusters consisting of 13 sites have been
used.

\begin{figure}[tbp]
  \centering \includegraphics{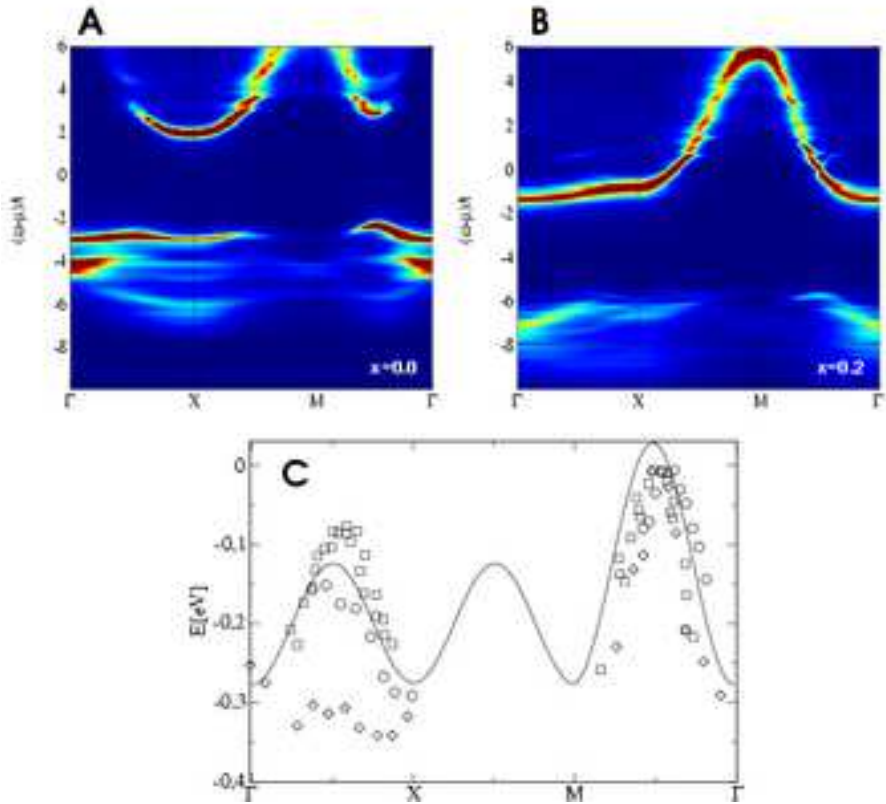}
  \caption{Spectral function of the $t-t'-U$ Hubbard model with $t'=-0.35t$
    and $U=8t$. Panel \textsf{A}: half filling. Panel \textsf{B}: overdoped system with
    $x=0.2$. Panel \textsf{C}: Detailed dispersion of the quasiparticle band
    of panel \textsf{A} for $t=0.5eV$. Symbols represent the experimentally
    determined dispersion for Sr$_{2}$CuO$_2$Cl$_2$
    (squares \cite{du.le.01}, circles \cite{la.vo.97}, diamonds \cite{we.sh.95}).
    Both spectra where obtained by a CPT calculation using a 10 site
    cluster.}
  \label{fig:1band-ntpype-akw-combine-2}
\end{figure}

Figure \ref{fig:1band-ntpype-akw-combine-2} displays the spectral function
$A(\mathbf{k},\omega)$ of the half-filled ($x=0.0$, panel \textsf{A}) and
overdoped $t-t'-U$ Hubbard model ($x=0.2$, panel \textsf{B}).
The plots show $A(\mathbf{k},\omega)$ along the momenta $\Gamma=(0,0)
\rightarrow X=(\pi,0) \rightarrow M=(\pi,\pi) \rightarrow \Gamma=(0,0)$
through the Brillouin zone.
The half-filled system in panel \textsf{A} gives rise to a narrow quasiparticle band,
roughly between $\omega=-3t$ and $\omega=-2t$.
A more detailed plot is given in panel \textsf{C}.
One notices the characteristic parabolic dispersion close to
$\mathbf{k}=(\pi/2,0)$, $\mathbf{k}=(\pi,\pi/2)$ and
$\mathbf{k}=(\pi/2,\pi/2)$.
Assuming $t \approx 0.5 eV$, this dispersion is practically identical to the
ARPES data \cite{ar.ro.02,ro.ki.02,du.le.01,la.vo.97}.
The indirect single particle gap between $\mathbf{k}=(\pi,0)$ and
$\mathbf{k}=(\pi/2,\pi/2)$ is about $4t$, which is the maximum value still
compatible with
 experiments.
%

\long\def\new#1{#1}

\new{
At about $\omega=-3 t$ a feature with maximal
spectral weight close to $\mathbf{k}=(0,0)$ can be observed.  The same
spectral feature was already observed 
in early QMC simulations of the single-band
$t-U$ Hubbard model~\cite{pr.ha.95},
 exact diagonalizations of the
t-J model\cite{ed.oh.97.p} and approximate perturbative methods such
as the self-consistent Born approximation (SCBA).
Its spectral weight was mostly perceived as coherent,
 i. e. corresponding to the
``coherent'' motion of a ``spin-bag'' like quasiparticle.
This is also supported by
 more recent QMC simulations\cite{gr.ed.00}, V-CPT calculations
\cite{da.ai.04} of
the $t-U$ Hubbard  model, and analytical considerations~\cite{do.za.99u}.
}

We now discuss 
the spectral function for the overdoped ($x=0.2$) system, which is 
 plotted in 
panel \textsf{B}.
Here,
one finds a metallic quasiparticle band with a flat dispersion just below the
Fermi level at $\mathbf{k}=(\pi,0)$.
The band crosses the Fermi level close to $\mathbf{k}=(\pi/2,\pi/2)$ and
$\mathbf{k}=(\pi,\pi/2)$ and therefore creates a large Fermi surface closed
around $\mathbf{k}=(\pi,\pi)$ in the Brillouin zone.
The quasiparticle band shows almost the same dispersion as the tight-binding
($U=0$) model with the same parameterization.
Deep below the Fermi level,
between $\omega=-6t$ and $\omega=-10t$, we can see the traces
 of the lower Hubbard band observed at half filling, i.e. the area between
$\omega=-4t$ and $\omega=-6t$ in panel \textsf{A}.
\begin{figure}[htbp]
  \centering  \includegraphics[scale=0.60]{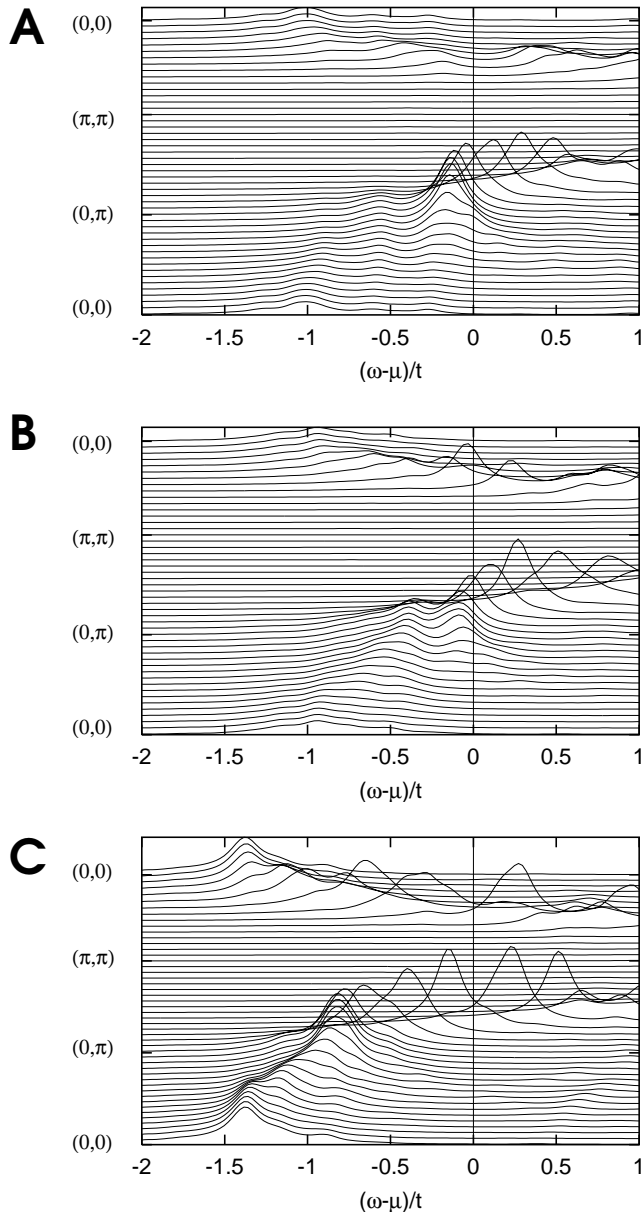}
   \caption{Details of the spectral function of the $t-t'-U$ 
     Hubbard model with $t'=-0.35t$ and $U=8t$ restricted to a
     narrower region around the Fermi level.  Panel \textsf{A}: x=0.077,
     obtained from a 13 site cluster.  Panel \textsf{B}: x=0.091, obtained
     from a 11 site cluster.  Panel \textsf{C}: x=0.200, obtained from a 10
     site cluster.}
  \label{fig:Line_Plots}
\end{figure}

In
Fig.~\ref{fig:Line_Plots} we plot a detail of the spectral function of the
Hubbard model in a more restricted region
 around the Fermi level for three doping values, namely
 $x=0.077$ (panel \textsf{A}), $x=0.091$ (panel \textsf{B}), and
$x=0.200$ (panel \textsf{C}).
For $x=0.077$, the Fermi level enters  into the upper Hubbard band with
only slight modifications of the spectral weight.
Most important, the arc around $\mathbf{k}=(\pi,0)$ in the upper Hubbard band of
the half-filled system is virtually unchanged and now forms an electron
pocket around $\mathbf{k}=(\pi,0)$, as can be seen in panel \textsf{A}.
However, for this underdoped system, 
the Fermi level has not yet reached the bottom of the parabolic band around
 $\mathbf{k}=(\pi/2,\pi/2)$.
The effect of doping is not limited to
a rigid shift of the band structure with respect to the
half-filled situation. As a matter of fact,
some new spectral weight is created between $\omega=-1t$ and $\omega=-0.5t$ at
$\mathbf{k}=(0,0)$ and between $\omega=-0.5t$ and $\omega=0t$ at
$\mathbf{k}=(\pi/2,\pi/2)$. 
%
%
At higher doping, 
for $x=0.091$ (panel \textsf{B}), this new spectral weight becomes more
pronounced and starts producing  the branches at $\mathbf{k}=(0,0) \rightarrow (\pi,0)$
and $\mathbf{k}=(0,0) \rightarrow (\pi/2,\pi/2)$ below the Fermi
level, which are clearly observed in the 
overdoped system in panel \textsf{C}.
%


\begin{figure}[tbp]
  \centering \includegraphics{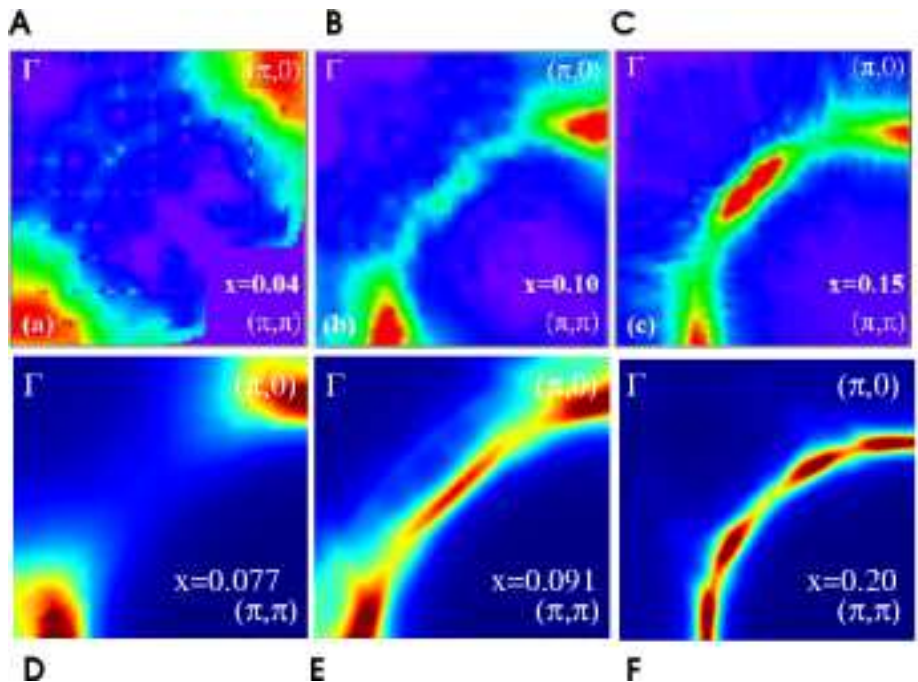}
  \caption{
Spectral weight at the Fermi energy, whose maximum is an indication of
the Fermi-surface contour for different doping levels.
The plots     \textsf{A} through \textsf{C} are 
obtained from ARPES experiments  (taken from \cite{ar.ro.02}), while
    \textsf{D} through \textsf{F} correspond to our CPT results, whose
    spectral functions are plotted in
    Fig. \ref{fig:Line_Plots}.  }
  \label{fig:Fermi_Surface}
\end{figure}

The results described above provide a deeper insight into the doping process of the
$t-t'-U$ Hubbard model, provided 
%
one assumes that the
parameters of the model do not change as a function of doping.
Although this assumption is widely believed to be appropriate for the doping range
considered here, there have been suggestions that 
 the on-site repulsion
$U$ is constant over a broader doping range.
%
%
The possibility of a doping-dependent on-site repulsion was considered recently on
the basis of a spin-density wave (SDW) mean field calculation of the $t-t'-t''-U$
Hubbard model to describe the experimental data of the electron-doped system NCCO
\cite{ku.ma.02,ar.ro.02}.
In contrast 
to the usual SDW calculation, where one self-consistently determines
the single particle gap $\Delta_\mathrm{mf}$ under the assumption that $U$ is
a fixed parameter, the authors of Ref. \cite{ku.ma.02} consider $U$
as a doping-dependent parameter fixed by the condition that at each
doping value
$\Delta_\mathrm{mf}$
corresponds to the
experimentally-determined pseudogap.
%
In their calculation, which we herewith refer to as KMLB,
the value of $U_\mathrm{eff}$ drops sharply upon doping from
$U_\mathrm{eff}=6t$ at half filling to $U_\mathrm{eff} \approx 3t$ at
$x\approx0.15$.
The success of this idea is supported by the fact that
the spectral function obtained 
from this procedure shows excellent agreement
with the doping evolution of the experimentally observed Fermi surface.
%
%
%
The whole scenario, however, is based upon the assumption that a varying
$U_\mathrm{eff}$ is indispensable for the reproduction of the experimental
results.

\begin{figure}[tbp]
  \centering \includegraphics{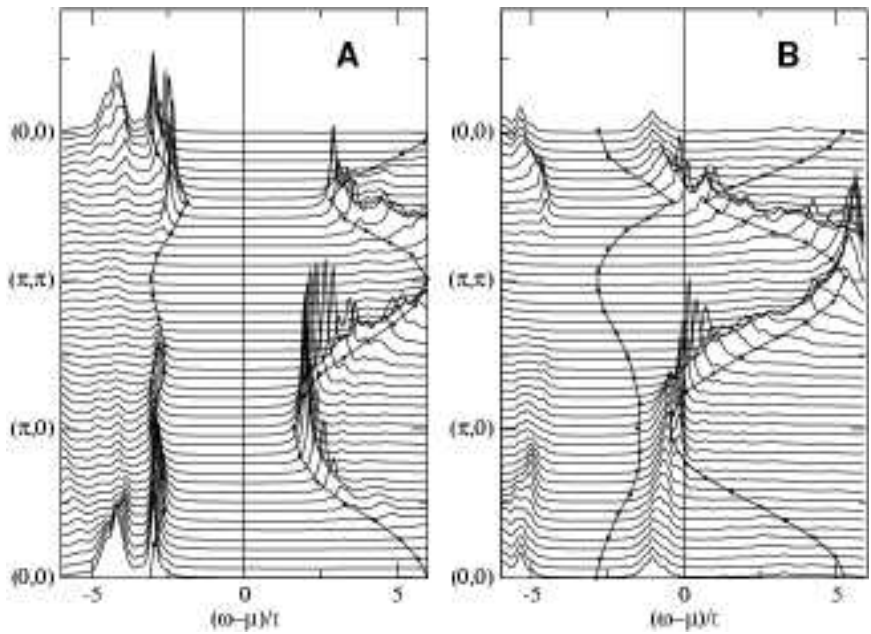}
  \caption{Direct comparison between CPT and KMLB results (see text). The KMLB dispersion
    is indicated by the solid line. 
In both cases, energies are given in units of $t=1$.
    A: half-filling; B: $x=0.10$ (CPT $x=0.091$}.
  \label{fig:1band-kusko-compare}
\end{figure}

For this reason, it is important to verify, whether a doping-dependent $U$ is necessary
in order to correctly reproduce the spectral features as a function of
doping, in particular the doping dependence of the Fermi surface. This
is conveniently represented experimentally by plotting an intensity
plot of the spectral weight at the Fermi energy $\omega=0$.
The upper row of Figure \ref{fig:Fermi_Surface} (panels \textsf{A} through
\textsf{C}) shows the ARPES data taken from \cite{ar.ro.02}, whereas the lower
row (panels \textsf{D} through \textsf{F}) plots the results obtained by
 in the CPT calculation.~\cite{end}  

Although we only used a minimal set of standard parameters and did not change
the parameterization (in particular $U$) as a function of
 doping, the CPT Fermi surface of the fully
correlated Hubbard model qualitatively reproduces the experimental result.
In particular, we observe electron pockets around $\mathbf{k}=(\pi,0)$
in the underdoped region, a FS patch at
$\mathbf{k}=(\pi/2,\pi/2)$ for about $x=0.10$ and a large FS centered
around 
$\mathbf{k}=(\pi,\pi)$ in the 
 overdoped case.
All these features are consistent with the experimental results 
 found in \cite{ar.ro.02} as well as in the Hartree-Fock calculation
 of Ref.~\cite{ku.ma.02}.
As a matter of fact,
 figure \ref{fig:Line_Plots} shows that the closing of the gap at
higher doping,
which is obtained ``by hand'' upon decreasing $U$ within the KMLB, 
comes about naturally within our numerical treatment of the Hubbard
model, in which electron correlations are treated more accurately.

Figure \ref{fig:1band-kusko-compare} directly compares the CPT and KMLB
quasiparticle dispersion at half filling and $x=0.10$ (CPT: $x=0.091$).
Notice that for the KMLB results only the dispersion is indicated
(solid line).
In order to have comparable energy scales, we reported the energies from
\cite{ku.ma.02} in unit of $t=1$.

At
 half filling (Panel \textsf{A}) both methods show almost identical
dispersions for the  lowest-energy excitations, since both methods have
been fitted
to the experimental results.
However,
as a SDW-type mean field method, the KMLB results cannot describe the full
lower Hubbard band and is exclusively fitted to the low-energy excitations.
Hence, the lower part of the spectrum between $\omega\approx-3t$ to
$\omega\approx-4t$ 
cannot be reproduced within
 this technique.

Both techniques describe the closing of the gap as a function of doping.
Although the CPT dispersion is much weaker and shows a much smaller gap at
$\mathbf{k}=(\pi/2,\pi/2)$, the qualitative development as a function
of doping is
very similar.
As discussed above,
there is additional information provided by CPT in contrast to KMLB,
concerning the high-binding energy part of the spectrum.
%
In particular, a remainder of the parabolic dispersion at
$\mathbf{k}=(\pi/2,\pi/2)$ is found at $\omega \approx -5t$.
This feature does not appear in the mean field calculation, but can be
clearly identified in the experimental data \cite{ar.ro.02}.



\section{Conclusion}
\label{sec:conclusion}
In conclusion, we have shown that the evolution of the Fermi surface of the
electron-doped cuprates is well described within the framework of the
 one-band Hubbard model with nearest- and next-nearest-neighbor hoppings.
In particular, we have provided indications that a doping-dependent
Hubbard repulsion $U$ is not necessary in order to describe the doping dependence of the
ARPES spectrum.
\section{Acknowledgements}
\label{sec:acknowledgements}
The authors would like to acknowledge support by the
DFG-Forschergruppe: Doping-dependence of phase transitions and
ordering phenomena in cuprate superconductors (FOR 538), and by the
Bavarian KONWHIR 
project CUHE. This research was supported in part by the National
Science Foundation under Grant No. PHY99-0794.
One of us (WH) would like to acknowledge the warm hospitality of the
Kavli Institute for Theoretical Physics in Santa Barbara, where part
of this work was concluded.

%
\bibliographystyle{prsty} 
\bibliography{references_database,refspaper1}

\begin{thebibliography}{10}
\bibitem{end}
Note that the doping levels are slightly different, since the available
  fillings in CPT are constrained by the cluster sizes.
\end{thebibliography}
%

\end{document}